\documentclass[
reprint,
 amsmath,amssymb,
 aps,
]{revtex4-2}

\usepackage{graphicx}
\usepackage{dcolumn}
\usepackage{bm}
\usepackage [english]{babel}
\usepackage [autostyle, english = american]{csquotes}
\usepackage[dvipsnames]{xcolor}

\begin{document}


\title{Unifying reciprocal and real space atomic dynamics in dilute gases}

\author{Jaeyun Moon}
\email{To whom correspondence should be addressed; E-mail: jaeyun.moon@ufl.edu}
 \affiliation{
Department of Mechanical and Aerospace Engineering,\\University of Florida, Gainesville, FL 32611, USA\\
Department of Materials Science and Engineering,\\University of Florida, Gainesville, FL, 32611, USA}

 

\date{\today}
\clearpage

\begin{abstract}
In solids, quanta of atomic vibrations are identified in reciprocal space by their frequency and wavevector as phonons. At the opposite end of the matter spectrum, dynamics of dilute gases is conventionally described in terms of atomic or molecular collisions and translations in real space and time. These two formalisms are apparently incompatible, leading to difficulties in understanding atomic dynamics in intermediate matter. In this work, we demonstrate that normal modes, often synonymously considered as phonons in solids, provide a microscopic description of various transport processes, including thermal conductivity, diffusion coefficient, and shear viscosity, in a prototypical dilute gas, argon. Our results bridge the conceptual divide between solid and gas phase descriptions and establish normal modes as a unifying framework for atomic dynamics well beyond crystalline solids.
\end{abstract}

\maketitle
\clearpage

Kinetic and scattering theories have laid the foundation for the microscopic explanation of a wide range of natural phenomena \cite{mcquarrie_statistical_1976, ashcroft_solid_1976, kittel_introduction_1976, landau_statistical_1958}. Early works by Drude and Lorentz modeled electrical conduction in metals in terms of classical electrons undergoing scattering with a characteristic relaxation time \cite{drude_zur_1900, lorentz_theory_1916}. Sommerfeld later incorporated quantum mechanics and Fermi–Dirac statistics into this framework, significantly improving its predictive power \cite{sommerfeld_zur_1928}. These developments were followed by Bloch’s formulation of electronic wavefunctions in periodic potentials, which established the foundation of band theory and modern electronic structure calculations \cite{bloch_uber_1929}. For dilute gases, pressure and temperature are described microscopically from atomic collisions and translations \cite{boyle_defence_1662, van_der_waals_over_1873, gay-lussac_recherches_1802, kronig_grundzuge_1856}. Beyond thermodynamics, Chapman and Enskog independently and successfully considered the Boltzmann transport equation to describe various transport properties of dilute gases such as shear viscosity and thermal conductivity \cite{chapman_mathematical_1939}.

In addition to these "real" particles, kinetic theory has also been monumental in describing thermodynamic and transport properties in solids in terms of "quasi-particles" \ in reciprocal space (frequency and wavevector) \cite{debye_zur_1912, lindsay_survey_2018, einstein_plancksche_1907, petit_recherches_1819}. In 1929, Peierls extended the Boltzmann transport equation to explain transport properties of interacting phonons in dielectric crystals \cite{peierls_zur_1929}. In the process, phonon interaction/scattering mechanisms including Umklapp and Normal processes were also proposed. For the last few decades, various other phonon interactions (e.g., phonon-defect and phonon-atomic tunneling) have been introduced with much success \cite{hanus_thermal_2021, allen_thermal_1989, simoncelli_unified_2019, sun_high_2020, deangelis_thermal_2018, phillips_two-level_1987, qian_phonon-engineered_2021}. Coupled with first-principles calculations, many normal mode calculations of transport properties have been successfully compared with measurements. \cite{togo_first_2015, lindsay_survey_2018}.

\begin{figure*}
	\centering
	\includegraphics[width=0.7\textwidth]{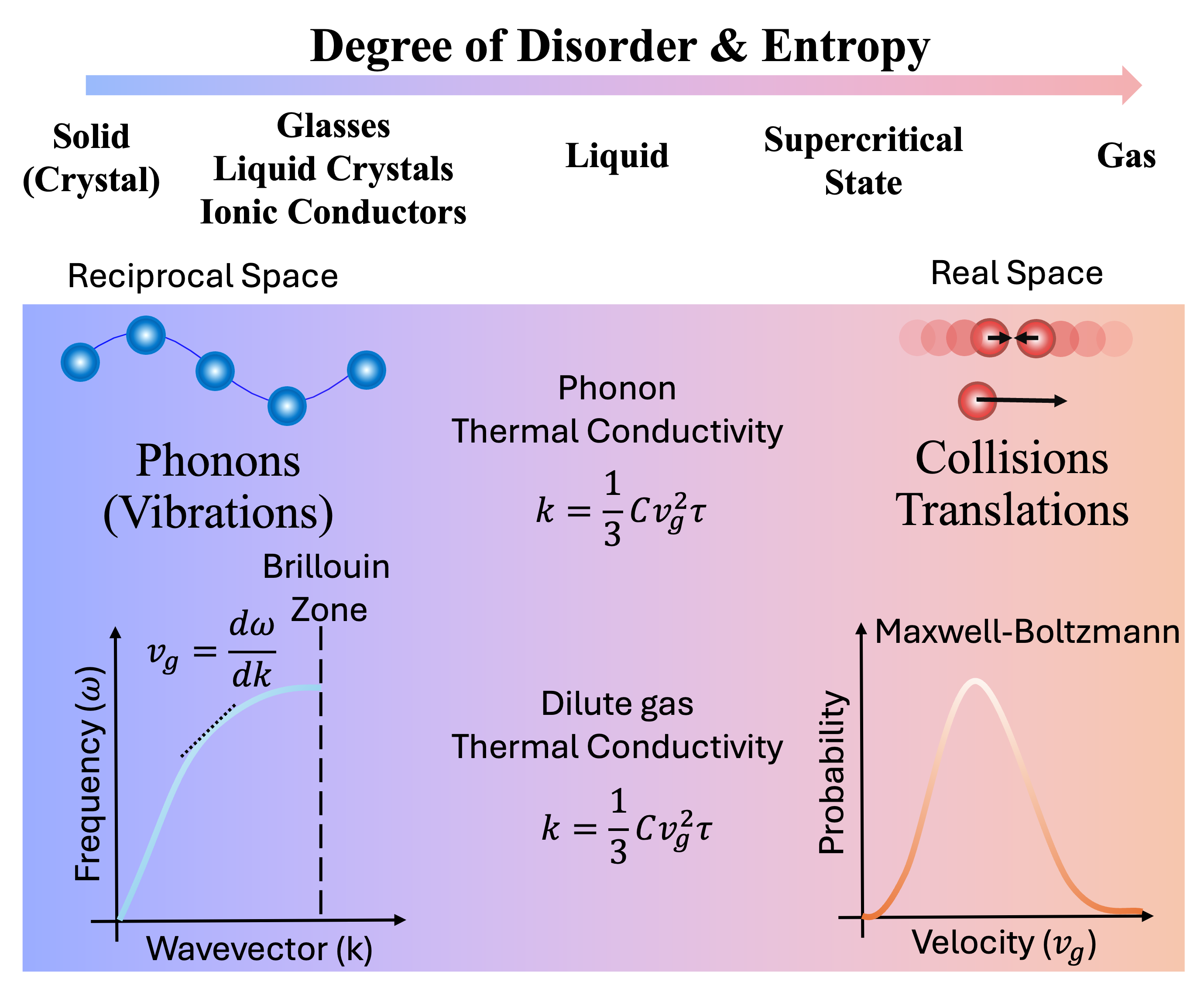}
	\caption{Schematic illustrating the fundamental divide in microscopic descriptions of thermal transport across phases arising from atomic motion. In crystalline solids, atomic motion is represented in reciprocal space through phonons characterized by wavevector and frequency. These phonons provide the foundation for describing thermodynamic and transport properties, including thermal conductivity. In contrast, dilute gases are described in real space, where atomic collisions and translations underpin macroscopic properties such as temperature, pressure, viscosity, and thermal conductivity. Despite these conceptually distinct microscopic pictures (reciprocal-space phonon dynamics versus real-space particle collisions), the resulting expressions for thermal conductivity take an identical kinetic form where $C$ is the specific heat, $v_g$ is group velocity, and $\tau$ is the relaxation time of appropriate heat carriers, as highlighted in the figure.}
	\label{fig:Overall}
\end{figure*}

Despite these remarkable advances in the microscopic understanding of macroscopic properties afforded by kinetic and scattering theories, a fundamental inconsistency remains. In gases, atomic degrees of freedom are described in terms of real-space dynamics, namely collisions and translational motion, whereas at the other end of the matter spectrum, solids are instead described using vibrations in the form of phonons in reciprocal-space (see Fig. \ref{fig:Overall}) \cite{moon_heat_2024}. These two physical pictures are apparently incompatible: real-space atomic collisions and translations cannot adequately describe thermodynamic and transport properties in solids, nor can vibrations be applied to gases. This dichotomy obscures a unified understanding of thermodynamic and transport phenomena in intermediate states of matter, including liquids, solid ionic conductors, and other systems that lie between the idealized limits of simple solids and dilute gases \cite{moon_disordered_2025}.

In our view, normal mode decomposition of atomic motion has a potential to fill this gap, as the number of normal modes in the system (equal to the number of atomic degrees of freedom) remains constant, regardless of the system of interest (whether solids or gases). Further, normal modes enable probing the curvature of local potential energy landscapes that govern thermodynamics and transport properties of materials. Normal modes therefore provide a complete and universal representation of atomic dynamics, even when they are not vibrational in the conventional sense. Recently, we demonstrated that normal modes in a dilute single element gas can depict real-atomic collision and translation motion dubbed collisons and translatons \cite{moon_atomic_2023}. Further, from a broader perspective, just as wave-particle duality lays the foundation for quantum mechanics, normal mode description of real atomic gas dynamics could also serve a rational basis.

In this work, we bridge the long-standing conceptual gap between reciprocal space and real space atomic dynamics descriptions through normal mode calculations and molecular dynamics simulations of a dilute gas (argon) as a prototypical gas for a wide range of temperatures up to 800 K at 1 bar. We characterize its temperature dependent transport properties such as thermal conductivity, diffusion coefficient, and shear viscosity through individual normal mode lifetimes. We show agreement within a few percent between our predictions and measurements from literature, demonstrating that normal mode bases describe atomic dynamics even in dilute gases. Our work, therefore, unifies the reciprocal and real space atomic dynamics descriptions of transport. 

\textit{Methods-} Dating back as early as Maxwell’s 1867 study of shear viscosity \cite{maxwell_iv_1867}, transport properties such as shear, bulk, and longitudinal viscosities, thermal conductivity, and diffusion coefficient have been considered in a generic form, $\chi = S\tau$ where $S$ represents a static, effective "modulus" \ and $\tau$ is a relaxation time or lifetime relevant to $\chi$ of interest \cite{iwashita_elementary_2013, jaiswal_atomic-scale_2015, malomuzh_maxwell_2019, ashwin_microscopic_2015, callaway_model_1959}. In dilute gases, all of these properties can be explained by the same underyling physics based on kinetic and scattering theory. For instance, thermal conductivity ($k$) and longitudinal viscosity ($\eta_l$) are simply related to shear viscosity ($\eta_s$) by 
\begin{equation}
    \eta_l = \frac{4}{3}\eta_s
\end{equation}
\begin{equation}
    k = \frac{15}{4}\frac{k_B}{m}\eta_s
    \label{Eq: k_eta}
\end{equation}
for a Lennard-Jones dilute gas where $k_B$ and $m$ are Boltzmann constant and atomic mass, respectively \cite{chapman_mathematical_1939}. Therefore, characterization of one dynamic property through understanding a corresponding lifetime can lead to general understanding of dynamics in dilute gases. Here, we focus on understanding thermal conductivity of argon gas through normal modes. 

\begin{figure}
	\centering
	\includegraphics[width=0.85\linewidth]{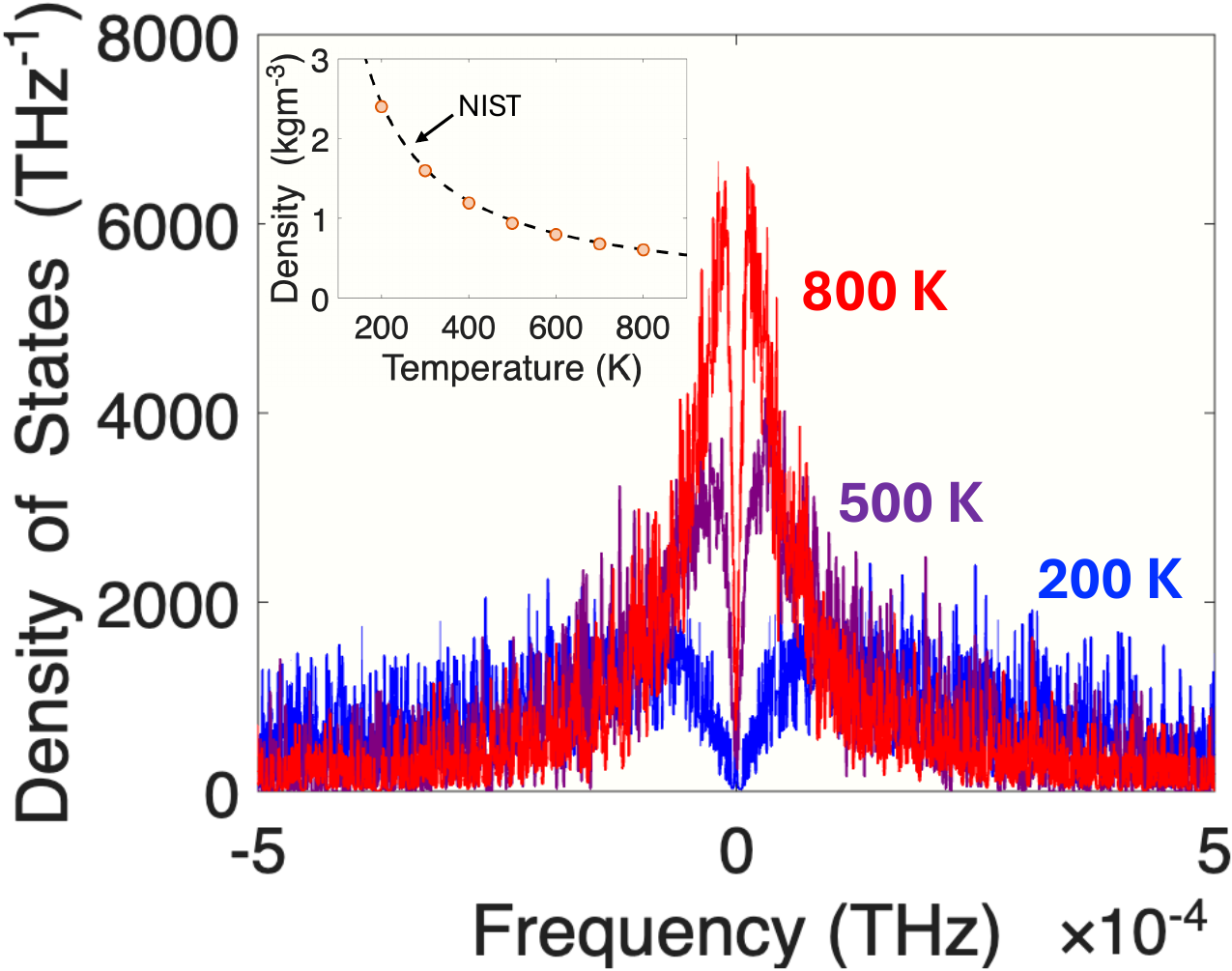}
	\caption{Representative temperature dependent normal mode densities of states of dilute argon gas at 200 K (blue curves), 500 K (purple curves), and 800 K (red curves). Nearly equal number of imaginary and real modes are observed at all temperatures, expected of a gas phase. Temperature dependent mass densities at 1 bar (brown circles) are shown in the inset along with NIST suggested measurement values (dashed curve) \cite{lemmon_thermophysical_2010}. Errorbars are smaller than the markers.}
	\label{fig:DOS}
\end{figure}

We first performed molecular dynamics (MD) simulations of argon (1372 atoms) using a Lennard-Jones potential \cite{jones_determination_1924, jones_determination_1924-1, rahman_molecular_1976} and LAMMPS \cite{plimpton_fast_1995} to generate equilibrated atomic structures from 200 K to 800 K with a 100 K interval under NPT (constant number of atoms, pressure, and temperature) ensemble at 1 bar. Total system volume is allowed to change. Resulting mass densities were less than 3 kgm\textsuperscript{-3}, about three orders of magnitude smaller than those of crystalline argon, as expected \cite{lemmon_thermophysical_2010} (shown in Fig. \ref{fig:DOS}).  Periodic boundary conditions are applied. For each temperature, argon gases were equilibrated for 10 ns to ensure thermal equilibrium under NPT, before data collection of another 300 ns, with a timestep of 10 fs under NVE (constant number of atoms, volume, and energy). Five different initial velocities were considered for ensemble averages.

At each temperature, atomic positions of a molecular dynamics simulation snapshot with a potential cutoff distance of half the simulation domain were used to diagonalize dynamical matrices ($\boldsymbol{D}$) at $\Gamma$ point to obtain normal mode eigenvalues ($\omega_n^2$) and eigenvectors ($\boldsymbol{e}_i$) (GULP \cite{gale_gulp_1997}) as
\begin{equation}
    \omega^2_i \cdot \boldsymbol{e}_i = \boldsymbol{D} \cdot \boldsymbol{e}_i.
\end{equation}
Dynamical matrices at $\Gamma$ point are related to the potential ($U$) by 
\begin{equation}
    D_{\alpha \beta, np} = \frac{1}{\sqrt{m_n m_p}} \frac{\partial^2 U_{np}}{\partial u_{\alpha,n} \partial u_{\beta,p}}
    \label{Eq: DM}
\end{equation}
where $m_n$ is the mass of atom $n$ and Greek subscripts denote Cartesian directions. Due to the instability of a snapshot structure, imaginary modes are expected \cite{keyes_instantaneous_1997, seeley_isobaric_1991, la_nave_instantaneous_2000, schirmacher_modeling_2022, baggioli_explaining_2021}.

Representative temperature dependent mode densities of states at 200, 500, and 800 K are shown in Fig. \ref{fig:DOS}. Imaginary modes are plotted as modes with negative frequencies. A majority of frequency magnitudes fall below the GHz range as opposed to the typical THz range for phonons in solids, due to extremely weak interatomic interactions. Frequency magnitudes become smaller with increase in temperature as interatomic distances become greater. Nevertheless, only three Goldstone modes with zero frequency are observed for all temperatures. In our prior work, we showed that eigenvalues in single element dilute gases do not represent square of vibrational frequencies but more generally a curvature of the potential energy surface \cite{moon_atomic_2023}. However, we still name the $x$ axis as frequency for historical reasons. We have approximately equal number of imaginary and real modes due to gas atoms having enough kinetic energy to overcome any of the hills and valleys of the potential energy landscape with equal probability as discussed in our prior work \cite{moon_normal_2024}.

With these temperature dependent normal modes, we next consider their lifetimes and connection to dynamic properties for dilute gases for the first time. In solids, normal mode lifetimes represent how energy autocorrelation of vibrational excitations or phonons decays over time. In an oscillator, its energy fluctuates between potential and kinetic energies in a periodic manner. As such, we obtain an oscillatory exponential decay for a mode kinetic energy autocorrelation as $\cos^2(\omega t)e^{-\frac{t}{\tau}}$ where $\omega$ is the radial vibrational frequency and $\tau$ is the mode lifetime \cite{mcgaughey_phonon_2006, moon_collective_2024, feng_quantum_2016, zhou_quantitatively_2015-2}. In contrast, for the single-component dilute gases considered here, we hypothesize a purely monotonic exponential decay without oscillations. This reflects the continuously evolving atomic environment and the absence of restoration that would return the system toward its initial configuration. Beyond characterizing energy decay, normal mode lifetimes in dilute gases can therefore be interpreted as a measure of how long the initial structural configuration and the associated forces remain correlated in time.  

\begin{figure}[h!]
	\centering
	\includegraphics[width=0.8\linewidth]{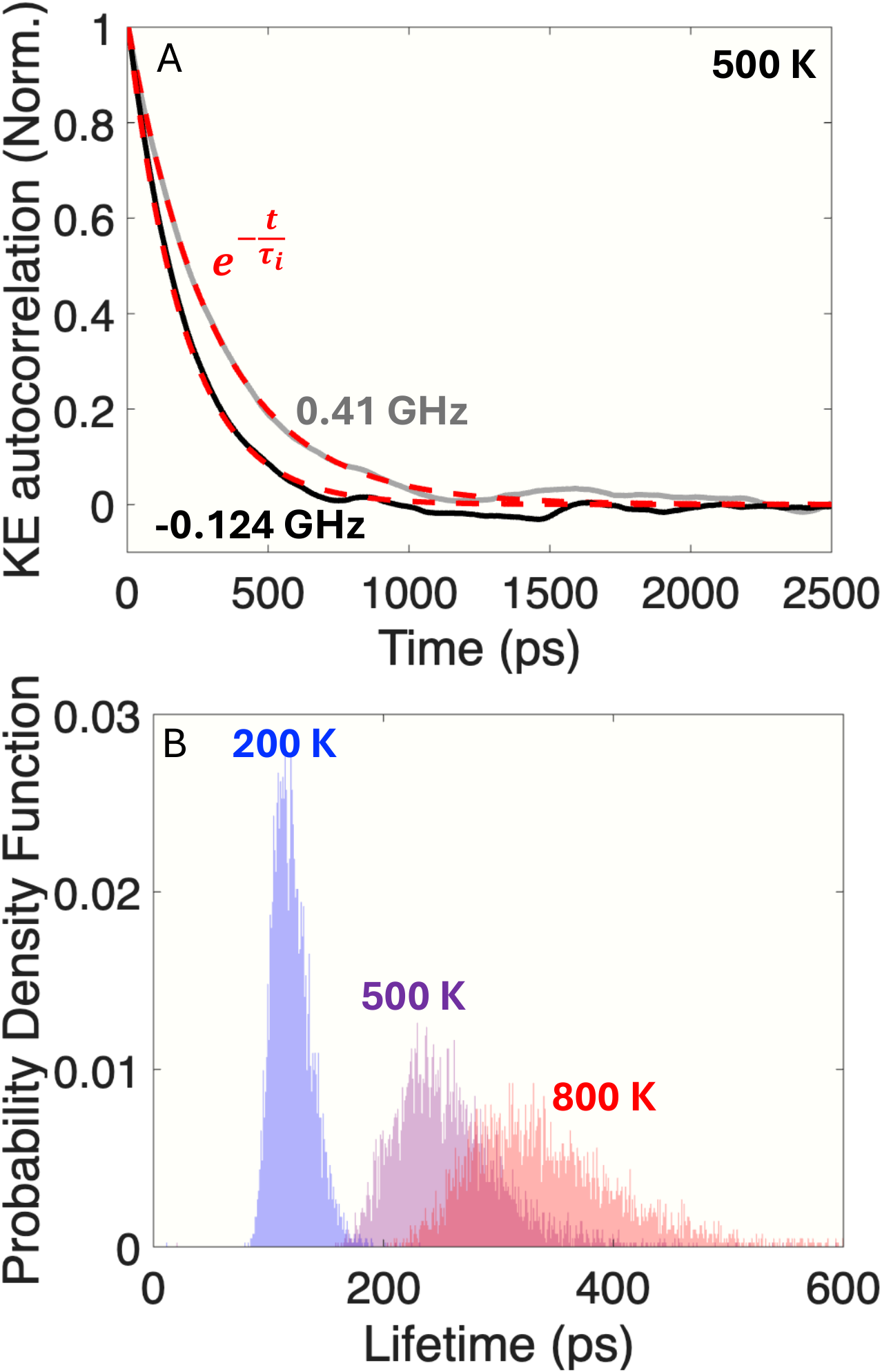}
	\caption{(A) Normalized, individual mode kinetic energy autocorrelation for two modes, -0.124 GHz (black curve) and 0.41 GHz (gray curve) for argon at 500 K. Red dashed curves are exponential decay fits where mode lifetimes are extracted. (B) Probability density function of normal mode lifetimes at 200 K (blue), 500 K (purple), and 800 K (red). Increase in lifetimes with temperature are shown.}
	\label{fig:mode_lifetime}
\end{figure}

 We calculated individual mode kinetic energy ($KE_i$) of our dilute argon systems as $\frac{1}{2}\dot{Q}_i(t)^2$ where $\dot{Q}_i(t)$ is the time derivative of normal mode coordinate for a mode $i$ and is given by
\begin{equation}
    \dot{Q}_i(t) = \sum_n m_n^{\frac{1}{2}}\boldsymbol{e}_{i,n} \cdot \boldsymbol{\dot{u}}_n(t).
\end{equation}
Sum is over all atoms, $\boldsymbol{e}_{i,n}$ is the eigenvector, and $\boldsymbol{\dot{u}}_n(t)$ is the atomic velocity \cite{moon_normal_2024-2, mcgaughey_predicting_2014}. Representative mode kinetic energy autocorrelations for two arbitrary modes are shown in Fig. \ref{fig:mode_lifetime}A for 500 K along with exponential decay fits. As anticipated, we observe no sinusoidal oscillations beyond some expected numerical error fluctuations at large times over $\sim$ 500 ps. Repeating this procedure for all individual modes, we obtain temperature dependent normal mode lifetime distributions. Normal mode lifetime probability density functions for 200, 500, and 800 K are depicted in Fig. \ref{fig:mode_lifetime}B. We see a clear temperature trend in the mode lifetime distribution. Unlike normal modes in simple crystals where lifetimes generally decrease with increase with temperature due to enhanced phonon-phonon interactions, mode lifetimes here increase instead due to a competing effect of thermal expansion and atomic velocities, consistent with kinetic theory of gases. Further, we did not observe qualitative differences in lifetime maginitudes or frequency dependence between real and imaginary modes.

Now we examine how these normal mode lifetimes (characterized above) are directly relevant to lifetimes of various dynamic properties such as thermal conductivity and diffusion coefficient. Thermal conductivity can be found by Green-Kubo formalism by 
\begin{equation}
    k = \frac{V}{3k_BT^2}\int_0^\infty \langle \boldsymbol{J}(0) \cdot \boldsymbol{J}(t) \rangle dt
    \label{Eq: GK}
\end{equation}
where $V$ is volume and angular brackets denote ensemble average \cite{kubo_fluctuation-dissipation_1966}. Heat current $\boldsymbol{J}(t)$ is given by 
\begin{equation}
    \boldsymbol{J}(t) = \frac{1}{V} \bigg[ \sum_n E_n(t)\boldsymbol{\dot{u}}_n(t) + \sum_{n<p} \big(\boldsymbol{F}_{np}\cdot \boldsymbol{\dot{u}}_p(t) \big)\boldsymbol{u}_{np} \bigg]
    \label{Eq: J}
\end{equation}
where $E_n(t)$ is atomic energy and $\boldsymbol{F}_{np}$ and $\boldsymbol{u}_{np}$ are the force and distance between atom $n$ and $p$, respectively. The second force interaction term, important for solids, is negligible for monatomic dilute gases. Integrand in GK thermal conductivity description is then dictated by atomic velocity autocorrelation functions due to nearly independent atomic velocities that lead to negligible velocity cross-correlations. Velocity autocorrelations, in turn, follow an exponential decay with time. Thermal conductivity for a monatomic dilute gas, after integration, can then be simply thought of as a static modulus multiplied by a heat current autocorrelation function lifetime or $\tau_{HCACF}$, similar to Maxwell relaxation time describing shear viscosity \cite{iwashita_elementary_2013}. 

\begin{figure}
	\centering
	\includegraphics[width=0.8\linewidth]{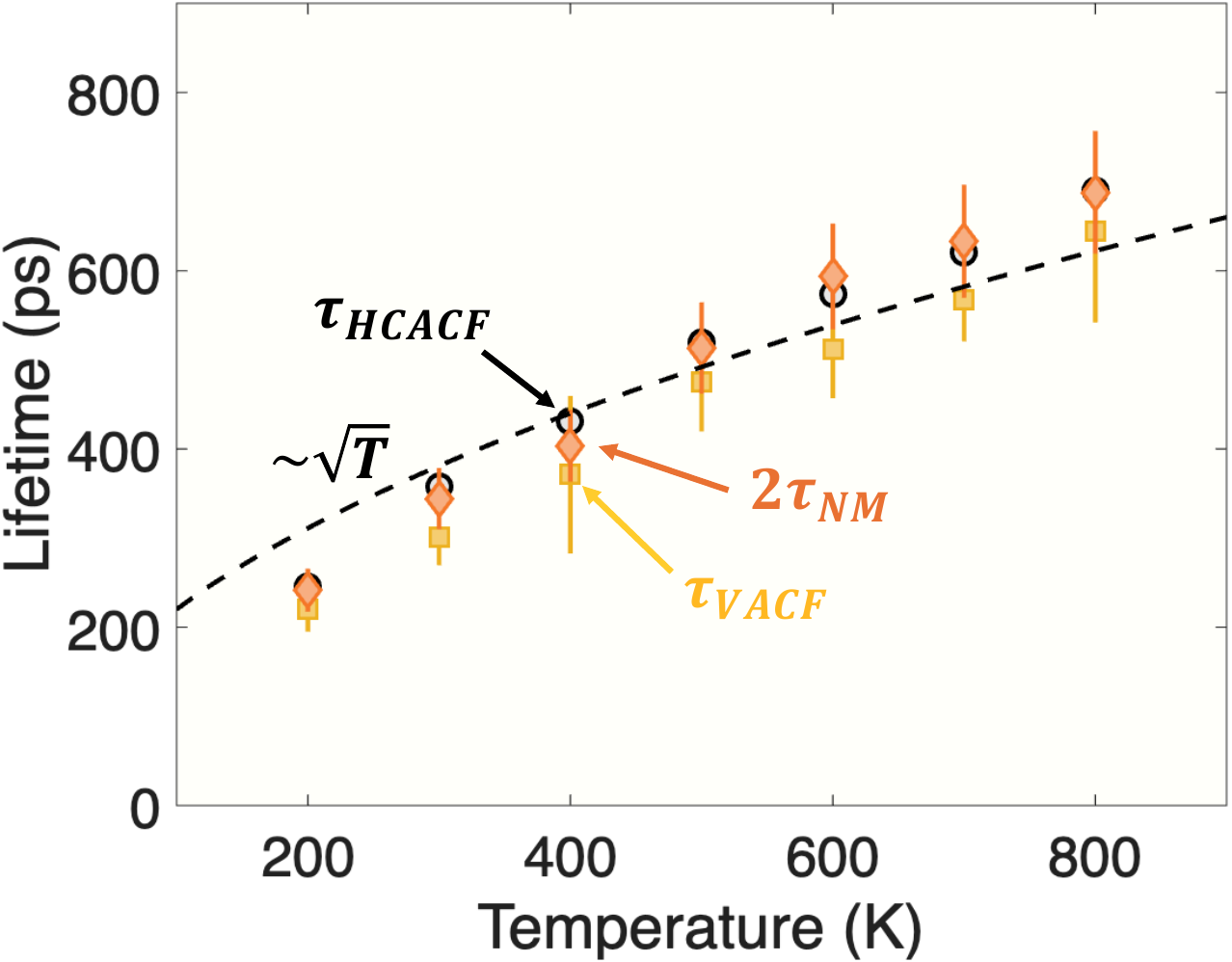}
	\caption{Temperature dependent lifetime comparisons. Black, orange, and yellow circles represent heat current autocorrelation function, mean normal mode, and velocity autocorrelation function lifetimes, respectively. Dashed curve shows a $\sqrt{T}$ dependence, expected for an ideal gas. Nearly overlapping lifetimes are seen for all temperatures studied here.}
	\label{fig:lifetime}
\end{figure}

Resulting temperature dependent $\tau_{HCACF}$ along with velocity autocorrelation lifetimes, $\tau_{VACF}$ characterizing diffusion coefficient and mean normal mode lifetimes $\tau_{NM}$ from Fig. \ref{fig:mode_lifetime} are compared together in Fig. \ref{fig:lifetime}. $\sqrt{T}$ scaling is also shown as reference, representing an expected temperature dependence for lifetimes of an ideal gas. A similar temperature dependence is demonstrated for all lifetimes considered here, especially at high temperatures though a slight deviation may be due to argon gas not being an ideal gas. We observe negligible effect of velocity cross-correlations as demonstrated by $\tau_{HCACF} \approx \tau_{VACF}$ (see Fig. S1 in the supplementary material). This means that thermal conductivity and diffusion coefficients which are characterized by $\tau_{VACF}$ are different only by the static modulus term for a dilute gas. In Fig. \ref{fig:lifetime}, a factor of two increase of $\tau_{NM}$ compared with the other two lifetimes is appropriate since normal mode lifetimes are found by autocorrelations of velocity squared whereas other lifetimes are autocorrelations of velocity to the first power. We observe a near-overlap between $\tau_{HCACF}$ and $2\tau_{NM}$ for all temperatures studied here. The near identity of $\tau_{HCACF}$, $\tau_{VACF}$, and $2\tau_{NM}$ demonstrates that transport lifetimes in dilute gases are not independent quantities, but manifestations of the same underlying relaxation process.

Next, we fully calculate the thermal conductivity using normal mode description of the GK formalism. Replacing the integrand with an exponential decay function and using appropriate thermal energy to simplify the energy and velocity variables yields the following   $\chi = S\tau$ relation as 
\begin{equation}
    k = \frac{5}{2}\frac{k_B^2TN}{mV}(2\tau_{NM}).
    \label{Eq: k_NM}
\end{equation}
This, in turn, is equivalent to the kinetic theory of gases as $k = \frac{1}{3}C_pv_g^2\tau$. With the density and temperature known, the only unknown is then the normal mode lifetimes. Shear viscosity can be subsequently found from Eq. \ref{Eq: k_eta}. These dynamic properties are compared against measurement values from NIST as shown in Fig. \ref{fig:properties} \cite{lemmon_thermophysical_2010}. An excellent agreement is shown, suggesting that normal modes, typically considered as a synonym for phonons in solids, fully represent atomic dynamics even in dilute gases like argon studied here. 

\textit{\begin{figure}[h!]
	\centering
	\includegraphics[width=0.98\linewidth]{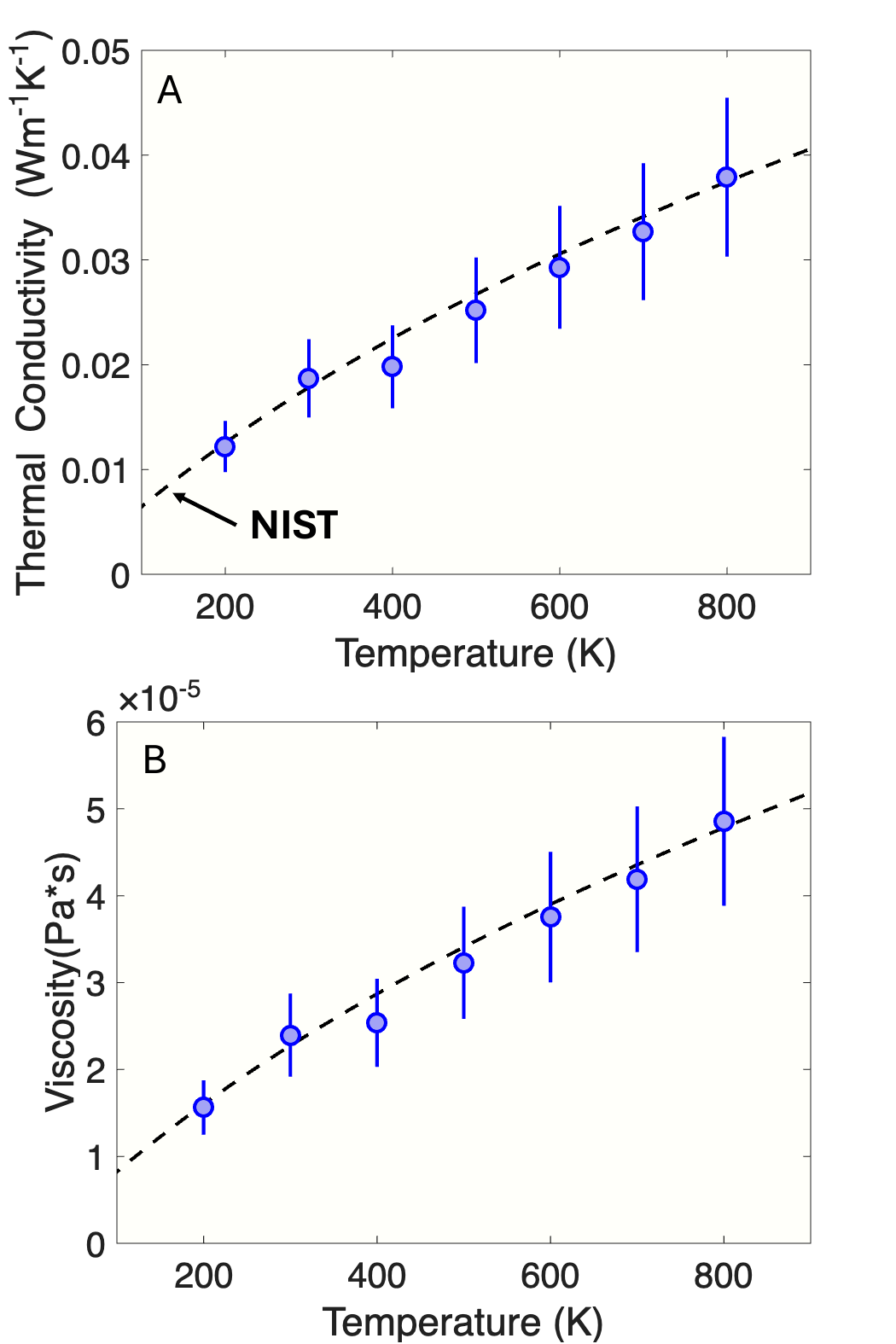}
	\caption{Isobaric thermal conductivity (A) and shear viscosity (B) from 200 K to 800 K at 1 bar for argon gas. Blue circles are from our normal mode calculations and dashed curves are measurement values from NIST \cite{lemmon_thermophysical_2010}. We observe an excellent agreement between our calculations and measurements within a few percent. }
	\label{fig:properties}
\end{figure}}

Conventional kinetic theory of gases describes transport in terms of discrete atomic or molecular collisions occurring in real space and time. However, in real dilute or dense gases, the notion of a well-defined discrete collision event becomes ambiguous: interatomic interactions continuously perturb particle trajectories within the range of the interaction potential, rather than producing abrupt, isolated scattering events (See. Fig. S2 in the supplementary material). As a result, accurately defining and estimating relevant mean free paths and lifetimes of atomic motion through real-space collision events in such gases remains an active area of research \cite{tsalikis_dynamics_2023, tsalikis_new_2024}. Without invoking mean free paths or collision cross sections, our work proposes an alternative framework in which lifetimes are characterized through individual normal modes defined in reciprocal, configurational space, the same conceptual framework used to describe atomic vibrations in solids.

We anticipate that our findings are relevant to multi-element dilute atomic gases, but a correction is needed for molecular gases with internal degrees of freedom or dense gases. However, these complexities may be described physically as vibrations, naturally included in the normal mode framework. The combination of vibrational modes, collisons, and translatons merits further investigations. 

Foundational transport theories such as phonons in solids or kinetic theories of real atoms in dilute gases historically began by choosing a specifically classified phase. However, the incompatibility of these theories leads to difficulties in microscopically understanding other materials that do not fall into these classifications. Here, normal mode calculations and molecular dynamics simulations of argon (as a representative dilute gas), show that dynamics of normal modes determines macroscopic transport properties in dilute gases, similar to that of solids. This work bridges the knowledge gap, and will impact the perception and description of atomic dynamics of matter beyond a specific phase.

\textit{Acknowledgement-} JM acknowledges the startup support from the University of Florida and is grateful for discussions with Drs. Lucas Lindsay and Xun Li and Profs. Takeshi Egami, Philip Allen, and Simon Thébaud. This work used the Advanced Cyberinfrastructure Coordination Ecosystem: Services \& Support (ACCESS) program which is supported by National Science Foundation grants \#2138259, \#2138286, \#2138307, \#2137603, and \#2138296 (Allocation MAT240051). 

\textit{Data Availability-} All data supporting the findings of this study are available within the article.

\bibliography{apssamp}
\end{document}